\documentclass[reprint,superscriptaddress,
aip,
amsmath,amssymb]{revtex4-1}
\usepackage{graphicx}
\usepackage{braket}
\usepackage{dcolumn}
\usepackage{bm}
\usepackage{color}
\usepackage{graphicx}
\usepackage{hyperref}
\usepackage{amsmath}
\usepackage{titlesec} 
\titlespacing\section{0pt}{10pt}{4pt}
\titlespacing\subsection{0pt}{10pt}{2pt}

\begin{document}
\title{Magnetic sensing at zero field with a single nitrogen-vacancy center}
\author{Till Lenz}
\affiliation{Helmholtz Institut Mainz, Johannes Gutenberg University Mainz, 55128 Mainz, Germany}
\author{Arne Wickenbrock}
\affiliation{Helmholtz Institut Mainz, Johannes Gutenberg University Mainz, 55128 Mainz, Germany}
\author{Fedor Jelezko}
\affiliation{Institute for Quantum Optics, Ulm University, Albert-Einstein-Allee 11, Ulm 89081, Germany}
\author{Gopalakrishnan Balasubramanian}\email{gopi.balasu@iom-leipzig.de}
\affiliation{Leibniz Institute of Surface Engineering (IOM), Permoserstraße 15, 04318 Leipzig}
\author{Dmitry Budker}\email{budker@uni-mainz.de}
\affiliation{Helmholtz Institut Mainz, Johannes Gutenberg University Mainz, 55128 Mainz, Germany}
\affiliation{Department of Physics, University of California, Berkeley, California 94720-7300, USA}
\date{\today}

\begin{abstract}
Single nitrogen-vacancy (NV) centers are widely used as nanoscale sensors for magnetic and electric fields, strain and temperature. Nanoscale magnetometry using NV centers allows for example to quantitatively measure local magnetic fields produced by vortices in superconductors, topological spin textures such as skyrmions, as well as to detect nuclear magnetic resonance signals. However one drawback when used as magnetic field sensor has been that an external bias field is required to perform magnetometry with NV centers. In this work we demonstrate a technique which allows access to a regime where no external bias field is needed. This enables new applications in which this bias field might disturb the system under investigation. Furthermore, we show that our technique is sensitive enough to detect spins outside of the diamond which enables nanoscale zero- to ultralow-field nuclear magnetic resonance. 
\end{abstract}

\maketitle

\section{Introduction}
Single nitrogen-vacancy (NV) centers are widely used as nanoscale sensors for magnetic and electric fields, strain and temperature\cite{Rondin2014,Dolde2008,Acosta2010} even under ambient conditions. Here, we demonstrate the application as magnetic field sensor without the use of a bias field i.e. at zero field.
Nanoscale magnetometry using NV centers allows one to quantitatively measure local magnetic fields produced by vortices in superconductors \cite{Thiel2016}, topological spin textures such as skyrmions \cite{Dovzhenko2018} and to perform nanoscale nuclear-magnetic-resonance spectroscopy down to the single-molecule level \cite{Lovchinsky2016}. Commonly an axial bias field (along the NV axis) is applied in order to lift the degeneracy of magnetic sublevels in the ground state and gain first-order sensitivity to magnetic fields. In some applications, such as for example zero- to ultralow-field (ZULF) nuclear magnetic resonance (NMR) spectroscopy \cite{Blanchard2016} or the study of ferromagnetic thin films \cite{Zazvorka2020}, the bias field significantly disturbs the sample under investigation perturbing or even disrupting the measurement. In this work, we demonstrate a method to extend the range of application of single NV center based magnetometry to this regime by using circularly polarized microwave (MW) fields similar to work performed on an NV center ensemble \cite{Zheng2019}. 

\begin{figure}[ht]
\centering
\includegraphics[width=\columnwidth]{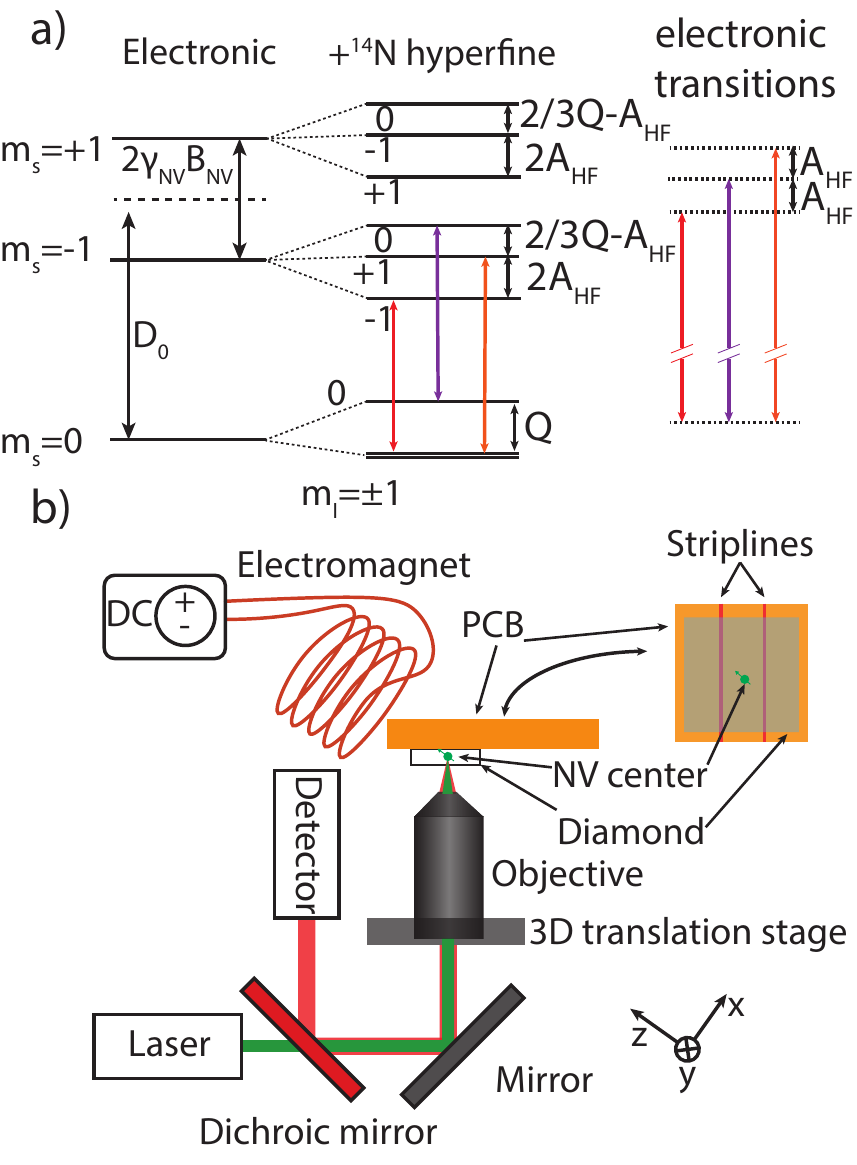}
\caption{\small{(a) Energy-level diagram of the NV center ground state without hyperfine interactions (left), which get split due to hyperfine interaction with the $^{14}$N host nucleus (center), and a representation of the resulting allowed electronic transitions  (right) b) Schematic of the experimental setup.}}
\label{setup}
\end{figure}

The NV center in diamond consists of a substitutional nitrogen atom and a neighbouring empty site in the diamond lattice. Sensing is performed with negatively charged NV centers, where an additional electron is obtained from a donor within the diamond. A simplified energy-level diagram of NV-ground state is shown in Fig.1a). The NV-center energy levels close to zero ambient field are described by the following Hamiltonian \cite{Kolbl2019,Steiner2010,Doherty2013}.
\begin{multline}
 \mathcal{H}/h=(D_{0}+\Pi_{z})S_{z}^{2}+\gamma_{NV}\boldsymbol{B}\cdot\boldsymbol{S}+\Pi_{x}(S_{y}^{2}-S_{x}^{2})
\\-\Pi_{y}(S_{x}S_{y}+S_{y}S_{x})+A_{HF}S_{z}I_{z}+Q[I_z^2-I(I+1)/3],
\label{ham}
\end{multline}
where $h$ is Planck's constant, $\gamma_{NV}=28$\,GHz/T the electron gyromagnetic ratio, $\boldsymbol{B}$ the magnetic field vector, $\boldsymbol{S}$ the electronic spin vector with its components $S_{x}, S_{y}, S_{z}$, respectively, and $D_{0}\approx 2.87$\,GHz the axial zero field splitting. $\boldsymbol{I}$ with components $I_{x}, I_{y}, I_{z}$ is the nuclear spin vector of the intrinsic $^{14}$N nucleus with the axial hyperfine coupling parameter $A_{HF}=-2.16$\,MHz and quadrupolar coupling $Q$. $\Pi_{x}, \Pi_{y}, \Pi_{z}$ describe the effective field acting on the NV center, which consists of strain and local electric field. The coordinate system is chosen such that the z axis is parallel to the NV axis.
For simplicity and without loss of generality we assume $\Pi_{z}=0$ for the scope of this work. In addition, note that the quadrupolar interaction between the $^{14}$N nucleus and the electron spin does not have an influence on the electronic transition frequencies as they are nuclear spin conserving (Fig.\ref{setup}a right). 

In most NV-magnetometry experiments a bias magnetic field is such that $|\gamma_{NV}B_z|>|\Pi_{x}|, |\Pi_{y}|, |A_{HF}|$ and $|B_x|,|B_y|\ll |D_0/\gamma_{NV}|$. In good-quality diamond samples $|A_{HF}|>|\Pi_{x}|, |\Pi_{y}|$ is also achieved. This leads to an energy-level structure shown in Fig.\,\ref{setup}a. The left side shows the simplified energy levels when omitting the last term in eq.\,\ref{ham}, which then allows for the measurement of the magnetic field by measuring the splitting between the $m_s=\pm1$ states. This spectrum with only two resonances increases in complexity, when the nuclear spin of the nitrogen host is taken into account (last term in eq.\,\ref{ham} and right side Fig.\,\ref{setup}a). Here it is shown for the case of a $^{14}$N nucleus with spin=1 (99.6$\%$ natural abundance). This means that there are three possible projections ($m_I=0,\pm1$) along the NV axis, which results in three resonances with a separation of $A_{HF}$.  The frequency difference of the central resonances of the two triplets is still $2\gamma_{NV}B_{z}$ (as shown on the right of Fig.\,1a). The  transition frequencies of the triplets cross at $\gamma_{NV}|B_z|=|A_{HF}|,|A_{HF}|/2,0$. At $B_z=0$ all three transition frequencies of the two sets cross (see also Fig.\,\ref{fieldscans}a).
In addition, due to the effective fields $\Pi_{x}, \Pi_{y}$, $S_z$ does not remain an eigenstate for the $\ket{m_I=0}$ states and a level anticrossing occurs. Both, crossing and anticrossings occurring at $B_z=0$ result in difficulties for magnetometry at $|B_z|<\Delta\nu/(2\gamma_{NV})$ where $\Delta\nu$ is the observed linewidth. In this regime, shifts of the transition frequencies which are proportional to magnetic field changes can not be resolved anymore or as for the case of an anticrossing (of states with equal $I_z$) the proportionality to magnetic is lost. In this work, we present a robust solution to this issue by the use of circularly polarized microwaves in the magnetometry protocol. Circularly polarized MWs allow us to exclusively drive transitions with $\Delta m_s=\pm1$. This leads to a single set of three resonances without crossings and therefore a linear relation to magnetic field even at zero magnetic field. 

For a single NV center  with the $^{14}$N nucleus in a fixed eigenstate, one would expect a single resonance for each of the two $\Delta m_s=\pm1$ transitions. However, two distinct sets of three resonances are observed as a result of the fact that the $^{14}$N spin flips several times within our measurement. The $^{14}$N basically acts as a source for a local bias field for the NV-electron system, where the value of the bias field depends on the nitrogen nuclear spin state. Due to the fact that the longitudinal relaxation time (T$_1$) for the $^{14}$N spin at room temperature is way below 10\,ms \cite{Neumann2010}, within a single measurement sequence (here: ODMR or Ramsey), the nuclear spin remains fixed, whereas over the time of the whole measurement one already observes an averaging over all three orientations of the nuclear spin, resulting in the sets of three resonances. We remark that, if one were able to fix the $^{14}$N nuclear spin in one of the $m_I=\pm1$ states, circularly polarized MW fields would have been unnecessary since the local bias would lift the degeneracy. Since this is not the case, we use circularly polarized MW fields to overcome this issue.

\section{Experimental setup}
In Fig.\,\ref{setup}b a schematic of our experimental setup is shown. It consists of a home-built confocal microscope with a high numerical aperture oil-immersion objective (Olympus UPLSAPO 60XO). The diamond sample containing single NV centers is mounted on a printed circuit board (PCB) which is used to apply circularly polarized MW fields to drive the transitions in the NV-center ground state. The design of the PCB follows earlier works \cite{mrozek2015,Zheng2019} and mainly consists of two parallel striplines (see Fig.\,\ref{setup}b).
The microwave pulses for spin manipulation were synthesized with an Arbitrary Waveform Generator (AWG; Tektronix AWG70002A), that operates at 50 GS/s. The microwave pulses (with adjustable relative phase) from the two output channels of the AWG  were independently amplified and fed-into the leads of the micro-circuit for producing circularly polarized microwave excitation. 
For these experiments we used a single native $^{14}$N-containing NV defect in an electronic grade diamond grown via chemical vapour deposition (CVD) from Element six. 
Furthermore, a coil is placed such that it exerts a controllable axial magnetic field to the NV center.

\section{Results}
\subsection{Characterization of the microwave polarization}
\begin{figure}[ht]
\centering
\includegraphics[width=\columnwidth]{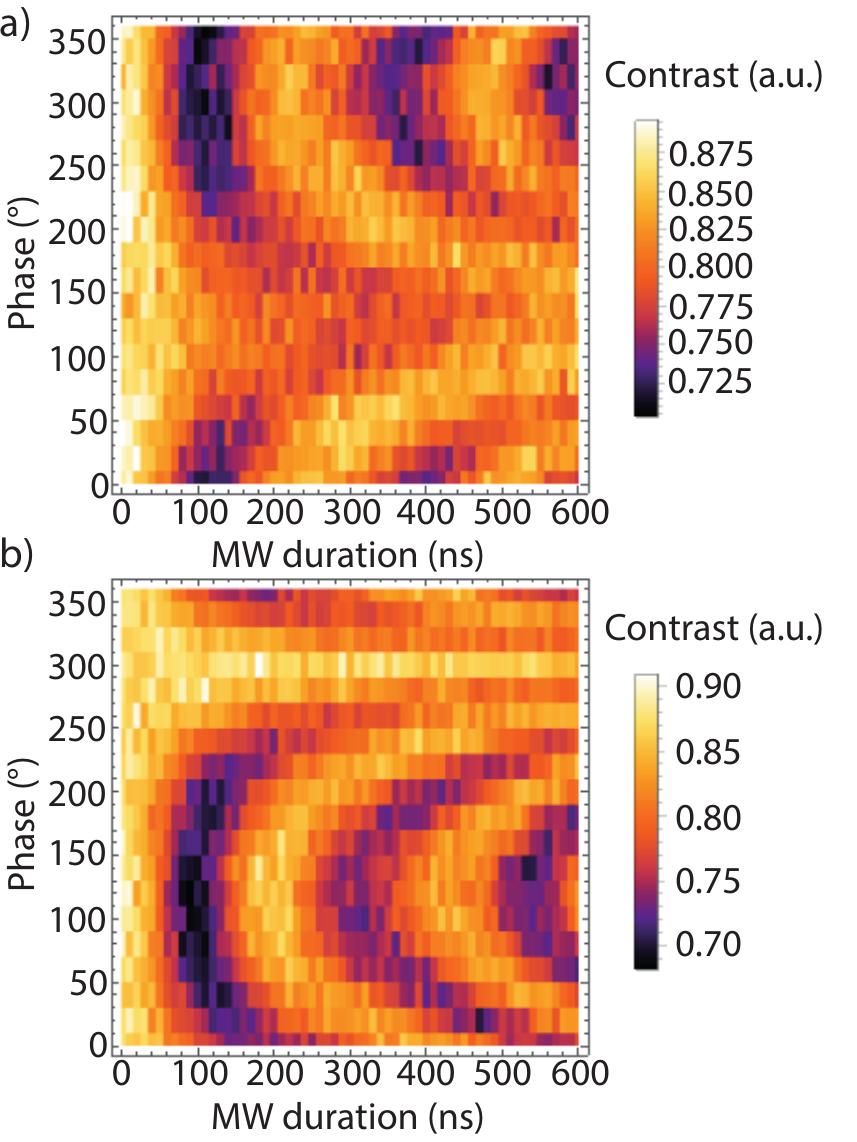}
\caption{\small{Rabi measurements performed to determine the driving strengths of the $\Delta m_s=-1$ (a) and $\Delta m_s=+1$ (b) transitions as a function of the relative phase of the MW fields. The measurements are performed at an axial bias field of $\approx450\,\mu$T.}}
\label{Rabi}
\end{figure}

First, a single NV center with proper alignment with respect to the microwave striplines needs to be identified. This was done by placing a permanent magnet at an angle of $\approx54^\circ$ with respect to the diamond surface and parallel to the MW striplines. Like this, the NV centers aligned with the with the magnetic field are also in the right orientation in order to apply circularly polarized MWs. After a single NV center aligned with the permanent magnet was identified ($B_z\approx450\,\mu$T), we performed Rabi-oscillation experiments on both the $\Delta m_s=+1$ and $\Delta m_s=-1$ transitions while sweeping the relative phase between the MWs applied to the striplines. Figure\,\ref{Rabi}a shows the $\Delta m_s=-1$ transition with a resonance frequency of 2859\,MHz and b the $\Delta m_s=+1$ transition with  a resonance frequency of 2882\,MHz . One can observe that both respective Rabi frequencies vary strongly upon change in relative phase between the striplines.  For the $\Delta m_s=+1$ transition the Rabi frequency varies between 4.3 and 1.45\,MHz and for $\Delta m_s=-1$ between 4.8 and $<0.8$\,MHz. 

From these values, we calculate the purity $p$ of the MW polarization defined as $p=\frac{\Omega_+ - \Omega_-}{\Omega_+ + \Omega_-}$. $\Omega_+$ and $\Omega_-$ are the respective Rabi frequencies of the $\Delta m_s=+1$ or $\Delta m_s=-1$ transitions. We obtain $p>71\%$ in this configuration. Note that the lower limit is given by the Rabi measurement obtained at a phase difference of $260^\circ$ while at $300^\circ$ the Rabi contrast in Fig.\,\ref{Rabi}b almost vanishes, which indicated that the maximum obtained purity of the circular polarization is significantly higher. The asymmetry and imperfect circularity of the MW polarization may arise from imperfect alignment of the striplines with respect to the NV center and was observed in earlier works with this stripline configuration \cite{mrozek2015}. Even though the polarization is not perfect, it is sufficient for our goals as we demonstrate in the following. \\

\subsection{Pulsed optically detected magnetic resonance}
\begin{figure}[ht]
\centering
\includegraphics[width=\columnwidth]{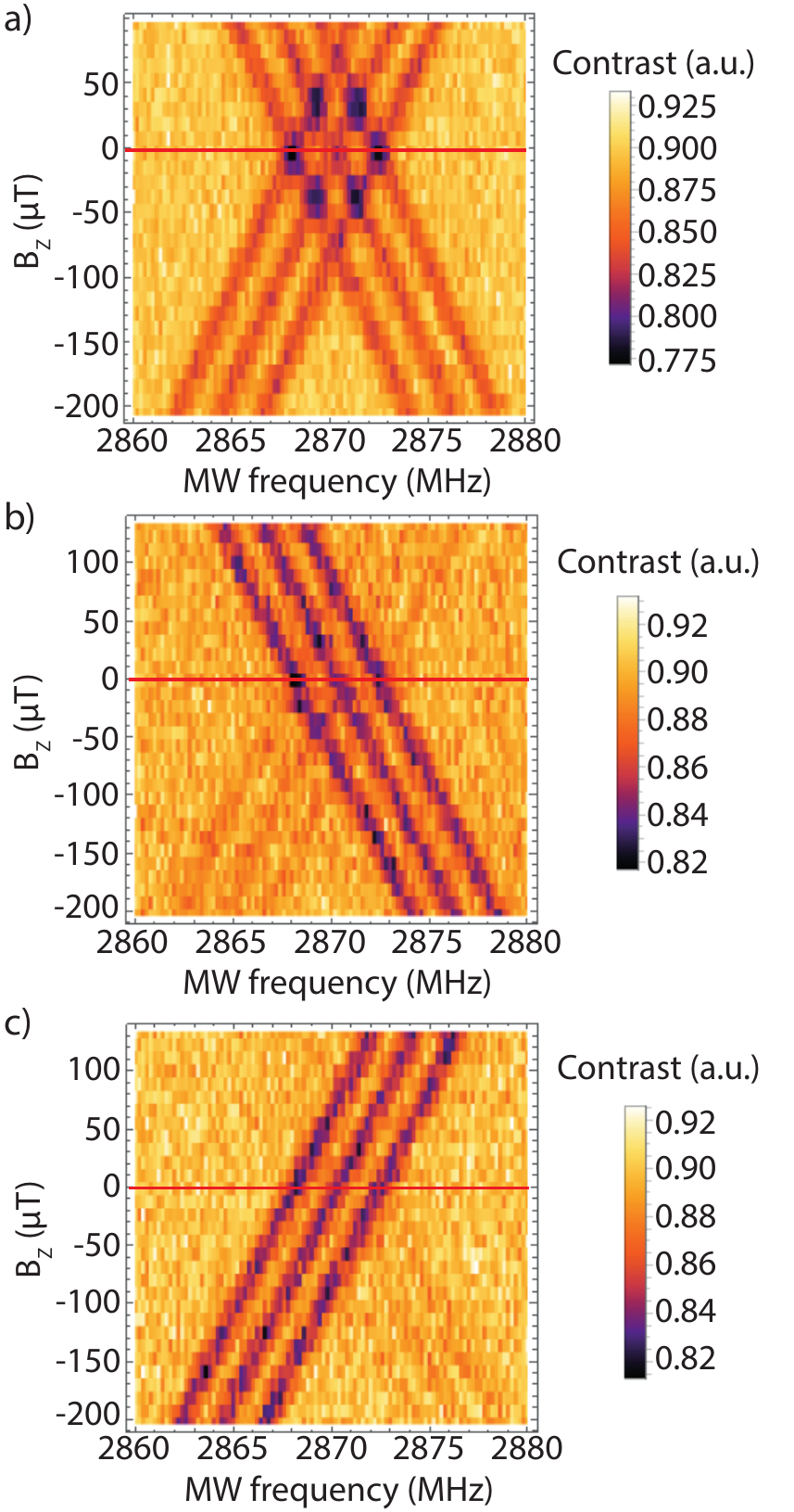}
\caption{\small{ODMR spectra for different MW polarizations while sweeping the axial magnetic field using an electromagnet. a) ODMR signal obtained with linearly polarized MW b) $\sigma^+$ polarization and c) $\sigma^-$ polarization.}}
\label{fieldscans}
\end{figure}

After characterizing the MW polarization the permanent magnet is replaced with an electromagnet using the same configuration with respect to the diamond and striplines (see Fig.\,\ref{setup}b).  Then axial field sweeps from $-200$ to 130\,$\mu$T are performed using the electromagnet. Figure\,\ref{fieldscans}a shows a density plot of the resulting pulsed optically detected magnetic resonance (ODMR) traces for linearly polarized microwaves. Note that the MW duration for each MW polarization is matched to a $\pi$-pulse when on resonance with the respective transition. At the bottom, i.e. at a field $|\gamma_{NV}B_z|>|A_{HF}|$ the two sets of three hyperfine resonances are well separated. When decreasing the applied bias field one can observe the above-mentioned line-crossings in the ODMR spectra. Especially at zero axial field (indicated by the red line) one observes three transition-frequency crossings. To be more precise, two crossings, while the central one is actually an avoided crossing due to the coupling of the two energy levels via electric fields and/or strain\cite{Mittiga2018}. In our case, this is characterized by lower contrast (also observable at $\approx\pm75\,\mu$T). In Fig.\,\ref{fieldscans}b, c the (anti-)crossings are not clearly observable anymore but when analyzing the individual ODMR we could still see the expected reduced contrast for the central resonance at zero field.

Figure \ref{fieldscans}b shows the case for the $\Delta m_s=+1$ transitions, while Fig.\,\ref{fieldscans}c shows a field scan for $\Delta m_s=-1$ transitions. The suppressed transitions are still slightly visible, but since they are strongly suppressed, they do not play a role when it comes to determining the transition frequencies of the strongly driven transition even at zero field. This shows already that magnetometry including advanced pulse sequences can also be applied at zero field if circularly polarized MW are used. 
\subsection{Ramsey sequence}
\begin{figure}[ht]
\centering
\includegraphics[width=\columnwidth]{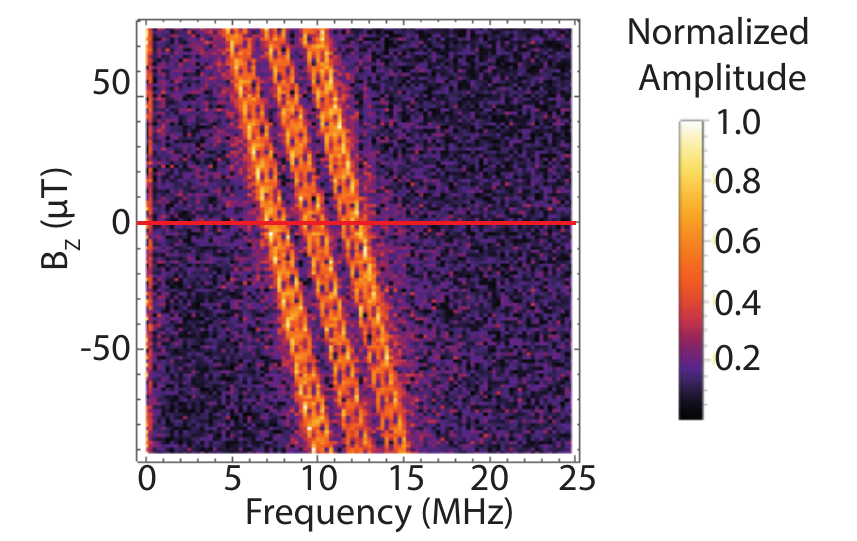}
\caption{\small{Fourier transform of the signal obtained from Ramsey experiments performed using $\sigma^+$ polarized MW while scanning the axial magnetic field, while the signal is effectively detuned by 10\,MHz due to the relative phase shifts of the two $\pi/2$ pulses. }}
\label{Ramsey}
\end{figure}

To enhance the magnetic sensitivity and in preparation for a future application of this technique in sensing low-frequency magnetic fields, we employ a Ramsey scheme\cite{Ramsey1950} . The frequency of the central anticrossing in Fig.\,\ref{fieldscans}a was chosen as driving frequency and the magnetic field was swept around zero axial field. In addition to that, the phase of the readout $\pi$/2 pulse was modified to implement a shift of the observed signals from DC to $\approx10$\,MHz. 
In Fig.\,\ref{Ramsey} the results of a Fourier transform of the initial Ramsey time trace is shown. One can clearly observe the same contrast over the full range of of axial magnetic fields, which again proves that we preserve the same magnetic sensitivity while crossing zero magnetic field. In addition to the $^{14}$N hyperfine splitting, we see hints of an underlying splitting induced by a nearby $^{13}$C nucleus in the Fourier-transformed signal, which is not fully resolved due to limited coherence time of our sensor spin. But this also demonstrates the robustness of this technique: even in the presence of additional couplings to different nearby nuclear spins, magnetometry around zero field can be pursued. 

To estimate the sensitivity of the technique, we determine the photon shot-noise limited sensitivity $\eta$ of a Ramsey measurement that is given by:
\begin{equation}
    \eta=\frac{\hbar}{g\,\mu_B}\frac{1}{C\sqrt{I_0 t_L/t_{seq}}\,T^{*}_{2}}
\end{equation}
with $I_0$ being the detected photoluminescence rate, $t_L$ the duration of the readout laser pulse, $t_{seq}$ the total sequence length and 
$T^*_2$ the coherence time of the NV center. Using common parameters used in our experiment, for this diamond sample where the coherence time is $\approx 3\,\mu$s, this results in a shot-noise limited sensitivity of $\approx350$\,nT/$\sqrt{\text{Hz}}$, which is a typical value for a diamond of this type \cite{Rondin2014}. This is sufficient to measure signals of, for example, proton spins cite{Muller2014} or other nuclear spins on the diamond surface or to study (anti-)ferromagnets using a shallow NV center and therefore this technique opens the field  to the study of the above mentioned systems using NV-center magnetometry at zero field. 

\section{Other possible approaches to magnetometry with NV centers at zero field}
As mentioned above, one could also use a static local bias field instead of a flipping bias ($^{14}$N,$^{15}$N) which would in turn avoid the need of specifically designed MW structures to produce circularly polarized MWs. A promising candidate for this is an axial $^{13}$C nucleus, which can, depending on the exact location, have flipping rates well below $10^{-4}$\,Hz or in an ideal case even zero. As a result both sets of resonances would be separated by the hyperfine splitting of the $^{13}$C nucleus and would change in frequency according to the Zeeman effect. A drawback to this method would be the need for an axial $^{13}$C which are usually randomly distributed with an abundance of 1.1$\%$ for natural-abundance diamonds. Moreover, the splitting produced by axial carbon is usually on the order of $\approx70$\,kHz which requires long NV coherence times to be resolved. Another option would be to initialize the nitrogen spin before each measurement into one of the $m_I=\pm1$ states, but common methods require a lifted of degeneracy between the $m_s$ states to function, which is not the case at zero field \cite{Pagliero2014}. For these reasons we focus on the approach with circular MW fields in this work but it is still worth mentioning other possible approaches.

\section{Summary and Outlook}
In summary, this paper presents a novel method to utilize NV centers as magnetic field sensors down to zero external magnetic field. This opens the road for applications in which magnetic fields perturb the system such as, for example, nuclear magnetic resonance and (anti-)ferromagnets. We also derived that the sensitivity of this technique is sufficient to probe statistically polarized nuclear spins on the diamond surface at zero field. In combination with high-resolution sensing techniques such as those developed for high-frequency sensing \cite{Schmitt2017,Glenn2018}, this technique may open new pathways towards nanoscale ZULF NMR for chemical analysis of liquids and solids down to the single molecule level\cite{Budker2019}. Further improvements on the driving strength would additionally make commonly used higher-order dynamical decoupling sequences  (such as XY8,CPMG,etc) accessible, which are used for AC magnetometry and therefore are particularly interesting for nuclear quadrupole resonance studies at zero field. 

\section{Acknowledgements}
This work is supported by the EU FET-OPEN Flagship Project ASTERIQS (action 820394), the German Federal Ministry of Education and Research (BMBF) within the Quantumtechnologien program (Grants No.FKZ 13N14439 and No. FKZ 13N15064), and the Dynamics and Topology Centre (TopDyn) funded by the State of Rhineland Palatinate.
GB would like to thank the Max Planck Society and MPI for Biophysical Chemistry for funding. We thank V. Kumar Kavatamane and John Blanchard for helpful advice. 
\bibliography{bibliography}
\end{document}